\documentclass{iaus}
\usepackage{graphicx}

\title[HI Fluff in the M\,81 Group] 
{Free--Floating HI Clouds in the M\,81 Group}

\author[Brinks, Walter \& Skillman]   
{Elias Brinks$^1$,
Fabian Walter$^2$ \and Evan D. Skillman$^3$}

\affiliation{$^1$Centre for Astrophysics Research, University of Hertfordshire, College Lane, Hatfield~~AL10~9AB, United Kingdom
\break email: E.Brinks@herts.ac.uk \\[\affilskip]
$^2$Max--Planck--Institut f\"ur Astronomie, K\"onigstuhl 17, 69117~Heidelberg, Germany \break email: walter@mpia.de  \\[\affilskip]
$^3$Dept.\ of Astronomy, Univ.\ of Minnesota, 116 Church St. SE, Minneapolis, MN 55455, USA\break email: skillman@astro.umn.edu}
\pubyear{2007}
\volume{244}  
\pagerange{143--149}
\date{?? and in revised form ??}
\jname{Dark Galaxies and Lost Baryons}
\editors{J. I. Davies \& M. J. Disney, eds.}
\begin{document}

\maketitle

\begin{abstract}
Recent VLA observations pointed at dwarf spheroidal (dSph) galaxies in the M\,81 group reveal a hitherto hidden population of extremely low mass ( $\sim10^5$\,M$_\odot$) HI clouds {\em with no obvious optical counterparts}. We have searched 10 fields in the M81 group totalling 2.2 square degree, both targeting known dwarf spheroidal galaxies and blank fields around the central triplet. Our observations show that the new population of low--mass HI clouds appears to be confined to a region toward the south--east of the central triplet (at distances of $\sim100$\,kpc from M\,81). Possible explanations for these free--floating HI clouds are that they are related to the dSphs found to the South--East of M\,81, that they belong to the galaxies of the M\,81 triplet (equivalent to HVCs), that they are of primordial nature and provide fresh, unenriched material falling into the M\,81 group, or that they are tidal debris from the 3--body interaction involving M\,81--M\,82--NGC\,3077. Based on circumstantial evidence, we currently favour the latter explanation.
\end{abstract}

\firstsection 
\section{Introduction} 

The nearby M\,81 group has been the subject of several surveys in the 21--cm line of atomic hydrogen (HI). These surveys can be divided into two broad categories: targeted and blind. \cite{appleton81} using the 76--m Lowell telescope and later \cite[Yun \etal\ (1994;]{yun94} \cite[see also Yun  \etal\ 2000)]{yun00} mapped  the large--scale HI distribution in the group, tracing the complex tidal tails and bridges resulting from the 3--body interaction between M\,81, M\,82, and NGC\,3077. Other targeted surveys, extending beyond the area covered by the triplet, have been those by \cite{huchtmeier98} and \cite{huchtmeier00} to determine the HI content of optically selected objects such as dwarf galaxies. Motivations for blind surveys have varied from i) searches for High Velocity Clouds (HVC) and Compact HVC analogues and  ii) deep surveys to push down the HI luminosity function, to iii) searching for ``Dark Galaxies", i.e., dark matter haloes which have not as yet turned (part of) their gas content into stars. The first blind survey which included the M\,81 group was that of \cite{lo79}. Recently, \cite{boyce01} have repeated this at improved resolution and sensitivity as part of HIJASS, the HI Jodrell All--Sky Survey.

\cite{boyce01} detected four known dIrr galaxies close to M\,81. They also detected an HI cloud apparently devoid of stars, HIJASS~J1021+6842. If anything else, this shows that blind HI surveys can still surprise and reveal objects which have been overlooked in one of the best studied and most surveyed nearby groups. \cite{walter05} present VLA follow--up observations of HIJASS~J1021+6842; \cite{kara07} report the possible detection of an optical counterpart in the form of faint H$\alpha$ emission which, if confirmed, would imply that although it would be one of the most extreme low surface brightness systems,  HIJASS~J1021+6842 wouldn't be a ``Dark Galaxy".

In what follows we will report on a project which started out with the intention to study the HI content of previously catalogued dSph galaxies in the M\,81 group. Deep observations were proposed to detect HI {\em in} these systems and {\em outside} the optical bodies in order to derive how much gas is associated with the galaxies themselves and how much is found in their immediate neighbourhood. We were interested in the latter to set constraints on the mechanisms that might be at work to remove gas from low--mass dwarfs. In the course of the project, however, we detected HI well beyond several of our targets, prompting us to change from targeted observations to blind survey mode.

\begin{table}
  \begin{center}
  \caption{Summary of  targets for the VLA D--array observations}
  \label{tab-obs}
  \begin{tabular}{lcllcc}
  \hline
 
	name       & type  & RA (2000.0) & DEC (2000.0)& m$_{\mathrm B}$ & detection\\
	       &   & h~~~m~~~ s &\ \  $^{\odot}$~~~$^\prime$~~~$^{\prime\prime}$ & mag \\
 \hline

	KDG 61     &  sph  & 09 57 03.1 & +68 35 31   & 15.2  &  y  \\
	FM 1       &  sph  & 09 45 10.0 & +68 45 54   & 17.5  &  n  \\
	BK 5 N     &  sph  & 10 04 41.1 & +68 15 22   & 17.4  &  y  \\
	KDG 64     &  sph  & 10 07 01.9 & +67 49 39   & 15.5  &  y  \\
	KK 77      &  sph  & 09 50 10.5 & +67 30 24   & 16.3  &  n  \\
	HIJASS     &  ?    & 10 21 00.0 & +68 42 00   & ?     &  y  \\
	DDO 71     &  sph  & 10 05 06.4 & +66 33 32   & 15.9  &  n  \\
	DDO 78     &  sph  & 10 26 27.4 & +67 39 16   & 15.8  &  n  \\
	BK 6 N     &  sph  & 10 34 29.8 & +66 00 30   & 16.9  &  n  \\
	KKH 57     &  sph  & 10 00 15.9 & +63 11 06   & 17.9  &  y  \\
  \hline
   \end{tabular}
  \end{center}
\end{table}

\section{Observations and Results} 

\begin{figure}
 \includegraphics{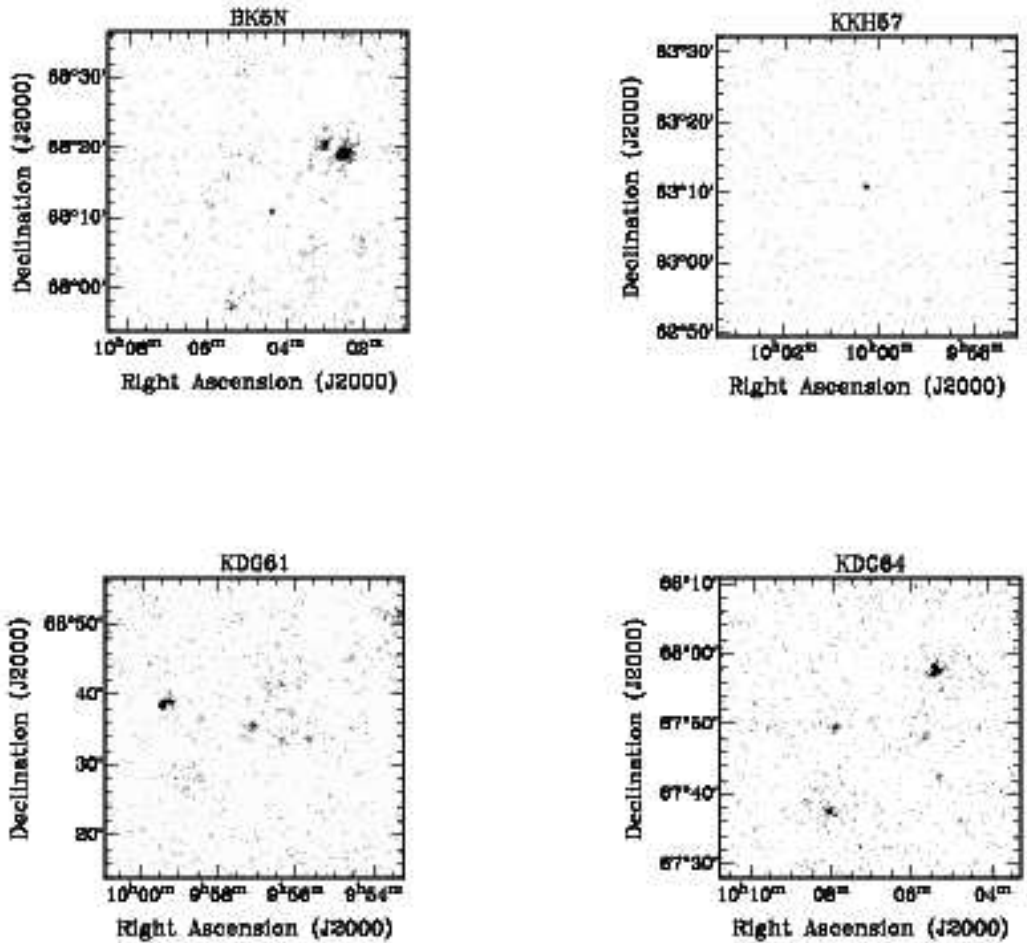}
  \caption{HI surface brightness maps of the four fields where HI emission was detected, BK5N, KKH\,57, KDG\,61, and KDG\,64. KKH\,57 is the only field where we detected HI associated with the target dSph; in the other three fields, free--floating HI is found whereas the target galaxies remain undetected down to a level of $9 \times 10^4$\,M$_\odot$ (based on a $3\sigma$ detection threshold over 3 channels).}\label{fig-obs}
\end{figure}

The observations reported here were made with the NRAO\footnote{The National Radio Astronomy Observatory is a facility of the National Science Foundation operated under cooperative agreement by Associated Universities, Inc.} Very Large Array (VLA) over an extended period. Originally 9 dSph galaxies, plus HIJASS\,J1021+6842 were observed in VLA D--array  in February 2003 (see Table~\ref{tab-obs}). In addition to mapping the source detected by \cite{boyce01}, HI was detected in the fields of BK5N, KKH\,57, KDG\,61, and KDG\,64. Follow--up C--array observations were obtained on these targets in April 2004. The resulting HI surface brightness maps of the combined C plus D--array observations are shown in Fig.\,\ref{fig-obs}. The maps were made using natural weighting for best sensitivity. They have an angular resolution of $35^{\prime\prime}$ which, at the assumed distance of the M\,81 group of 3.63\,Mpc \cite[(Freedman \etal\ 2001)]{freedman01}, corresponds to 0.6\,kpc. The rms noise is typically 0.6\,mJy\,beam$^{-1}$ or 0.3\,K at a velocity resolution of 5.2\,km\,s$^{-1}$. These single channel noise levels can be used to calculate a column density detection threshold. Assuming a signal to be genuine if it is detected at the $3\sigma$ level across 3 channels, we find a minimum detectable column density of $2.5 \times 10^{19}$\,cm$^{-2}$. Using the same criterion we arrive at a minimum detectable HI mass for an HI cloud filling the beam, at the distance of the M\,81 group, of $9 \times 10^4$\,M$_\odot$.

The results of our observations can be summarised as follows:

{\bf KKH\,57} We find a clear detection coinciding with the optical counterpart. The HI signal is only four channels wide at a systemic velocity of 203\,km\,s$^{-1}$; spatially the HI is unresolved. The low M$_\mathrm{HI}$/L suggests this is a transition galaxy, i.e., an object which combines properties of a dSph, such as a low luminosity and prominent old stellar population, with those of a dIrr, e.g., the presence of HI \cite[(see Mateo 1998, for a more detailed treatise]{mateo98} \cite[and Skillman \etal\ 2003, for examples)]{skillman03}. No other clouds are detected in this field.

{\bf KDG\,61} \cite{boyce01} claim that this is a clear case of a transition galaxy. Their beam, however, is $12^\prime$ large and they are limited by confusion due to HI coming from the triplet. KDG\,61 lies close to M\,81 and, even with our much higher resolution data, it is not trivial to separate emission from KDG\,61 or other field objects from M\,81 emission. We seem to find several HI clouds which are neither related to the SE tidal arm nor to KDG\,61 itself, at velocities ranging from $-45$ to $-15$\,km\,s$^{-1}$.

{\bf BK5N} Several HI clouds are detected with velocities from $-130$ to $-75$\,km\,s$^{-1}$, well offset from the optical counterpart (for which no optical radial velocity has been published). 

\begin{figure}
\begin{center}
 \includegraphics[width=11cm]{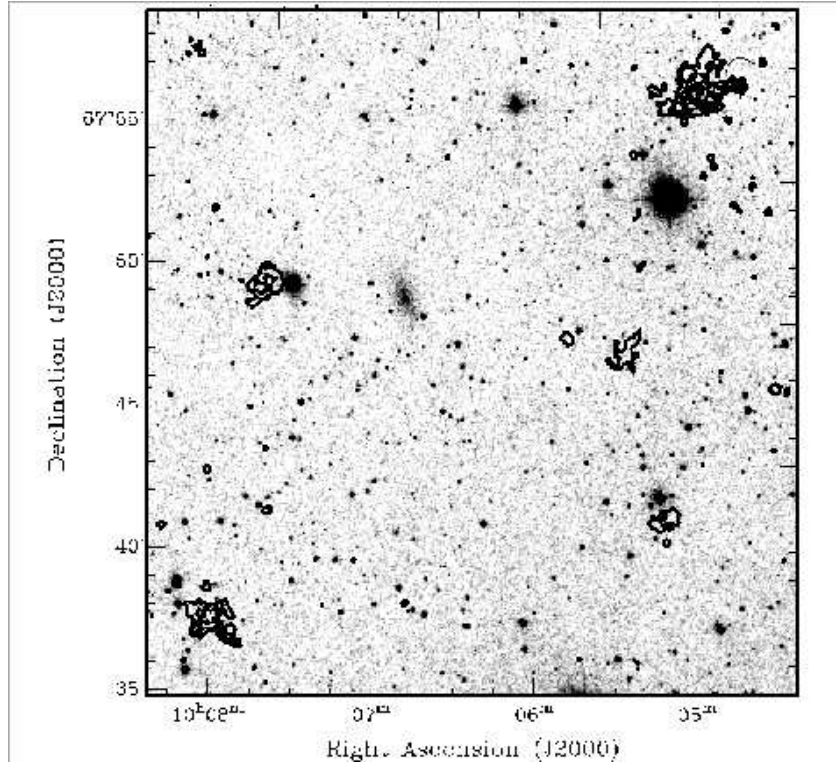}
\end{center}
  \caption{Optical image based on the {\em Digitized Sky Survey (DSS)} covering the field imaged in HI with the VLA centred on KDG\,64. The contours indicate the locations of half a dozen free--floating HI clouds. Note that no HI is associated with KDG\,64.}\label{kdg64}
\end{figure}

{\bf KDG\,64 = UGC\,5442} The optical galaxy lies at $-18 \pm 14$\,km\,s$^{-1}$ \cite[(Simien \& Prugniel 2002)]{simien02}. We find HI emission offset in position and velocity, with HI detected at velocities ranging from $-120$ to $-70$\,km\,s$^{-1}$. This is illustrated in Fig.\ \ref{kdg64} which shows the location of the HI clouds with respect to the optical galaxy.

In summary, except for KKH\,57, none of the dSph targets were detected, confirming that most of them are devoid of HI. However quite surprisingly, HI was detected in several of the fields centred on a number of our targets but not obviously associated with them, neither in position nor in velocity. This then, of course, begs the question if these clouds are somehow associated with the dSph galaxies, or if they are semi--randomly distributed within the M\,81 group. In order to investigate this, follow--up ``blind" HI observations, of a further six fields were obtained with the VLA in D--configuration (observations taken in June, July, and August 2004). HI was found in only one of these fields, bordering BK5N and KDG\,64 to the East. 

In total, about a dozen barely resolved HI clouds were detected. They seem to predominantly occupy a region to the South--East of the triplet, extending as far away as $\sim 100$\,kpc. Their radial velocities are in the range  $-130$ to $-70$\,km\,s$^{-1}$. Velocity dispersions are $\sim 8$\,km\,s$^{-1}$. HI masses are of order $10^5$\,M$_\odot$. Figure 2 shows a DSS image of the field mapped around KDG\,64 with some of the HI clouds indicated with contours. No optical counterparts are visible on the DSS. The M\,81 group has been surveyed several times in the optical down to levels considerably fainter than the DSS.  In fact, it is one of the best studied nearby groups \cite[(e.g., Froebrich \& Meusinger, 2000)]{froebrich00}. We can therefore be confident that these objects have very little starlight associated with them. 

\section{Discussion}\label{discuss}
So what are they and what is their origin? There are several possibilities. They could be:

\begin{enumerate}
\item primordial material which is infalling towards the M\,81 triplet;
\item High Velocity Cloud (HVC) analogues belonging to M\,81;
\item  material originally part of BK5N and KDG\,64 which has been expelled;
\item or tidal debris scattered into the SE quadrant.
\end{enumerate}

Taking each of these possibilities in turn, the first explanation, infall of primordial material, is rather {\em ad hoc}. Without a determination of the metallicity of the gas in these HI clouds, it will be difficult to rule out this hypothesis. However, if primordial material were still to be around at column and space densities found here, it should have been detected in other nearby groups as well. No similar objects have been seen in deep HI observations, reaching comparable HI detection limits (\cite[Pisano \etal\ 2004, 2007]{pisano04, pisano07}).

If they represent analogues to HVCs, one would expect their velocity centroid to coincide with that of their host galaxy, M\,81. Also, their distribution would be expected to be more isotropic rather than that found here. No clouds were found in any of the other targeted observations nor in the blind pointings which together covered areas in and around the triplet. The third option, that they represent material that has been expelled from BK5N and KDG\,64 is also hard to defend. Not only do the clouds seem uncorrelated spatially with both dwarfs, but in the one case where we have velocity information for the dwarf, the velocities of the clouds are offset substantially.

\begin{figure}
 \includegraphics[width=14cm]{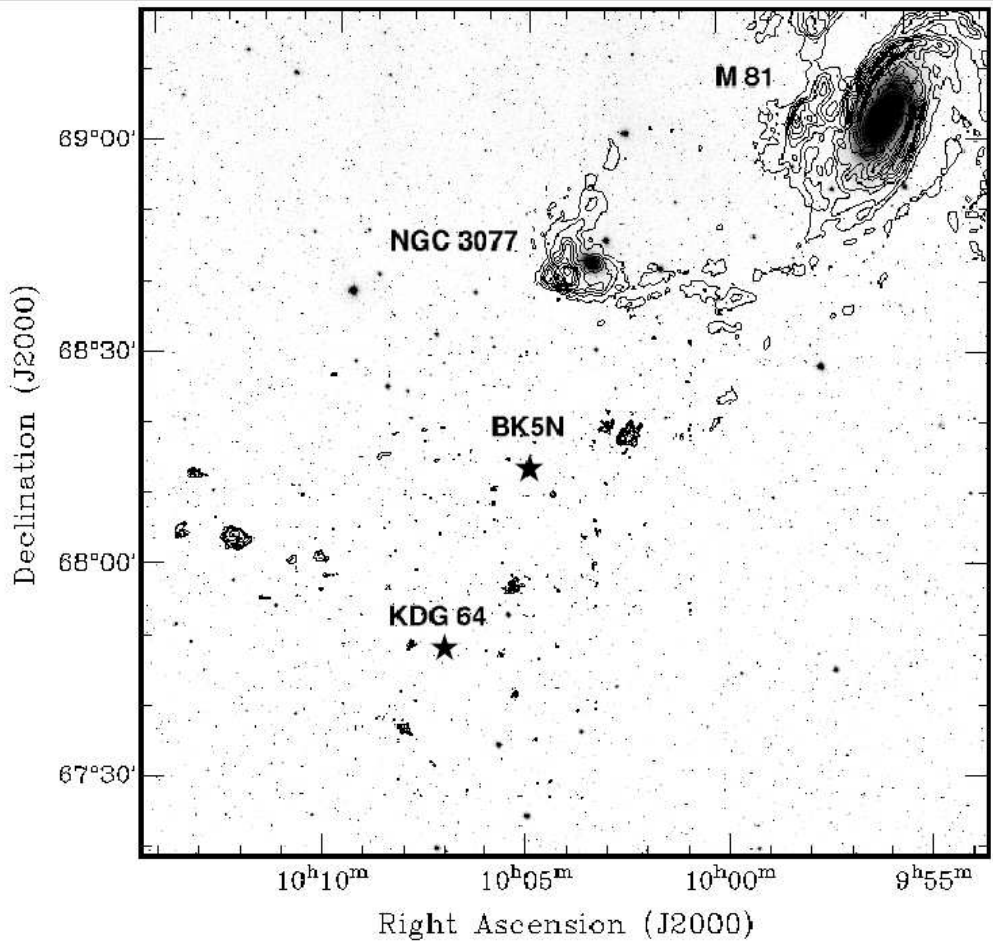}
  \caption{Optical image taken from the DSS of a $2^\circ \times 2^\circ$ field to the South--East of M\,81. Overlaid are contours of the HI distribution as measured by \cite{yun94}, showing the tidal tail connecting M\,81 with NGC\,3077, and a ``spur'' pointing to the South--East. In addition we plot as contours the HI clouds detected in the VLA observations reported here that are seen in the direction of BK5N and KDG\,64, plus the HI clouds encountered in VLA D--array observations in one of the ``blind" pointings located due East of the two dwarf galaxies. The approximate locations of BK5N and KDG\,64 are indicated with stars.}\label{m81}
\end{figure}

This leaves us with the last option, the clouds being tidal debris. Inspection of the HI observations by \cite{appleton81} shows that the HI clouds fall along the extension of their ``feature VII", but beyond the area covered by either their survey or the VLA mosaic by \cite{yun94}. Feature VII is prominent in the velocity range from $-130$ to $-100$\,km\,s$^{-1}$, in broad agreement with the clouds found near BK5N and KDG\,64.  The VLA map by Yun \etal\  resolves feature VII, showing seemingly a spur branching off from the tidal bridge connecting NGC\,3077 with M\,81, which points roughly in the direction of the dwarf spheroidals. This is illustrated in Fig.\ \ref{m81} which shows Yun's HI surface brightness map as contours overlaid on an optical image. The numerical simulation published by \cite{yun97} shows material from M\,81 being dragged out by the passage of NGC\,3077 and spread out towards the South--East. It therefore seems likely that the HI clouds found there are, in fact, tidal debris. They could be the neutral density peaks of a more dilute sheet of gas which is mostly ionised by the extragalactic radiation field, as a result of the column density being in general close to the value below which HI will become fully ionised. Peak column densities fall well below the canonical $\sim 10^{21}$\,cm$^{-2}$ threshold for star formation, explaining the lack of an optical counterpart.

\section{Conclusions}\label{sec:concl}

In a search for HI in and around the lowest--mass dwarf spheroidal (dSph) galaxies in the M\,81 group, we have discovered an unexpected population of HI clouds with masses of order $10^5$\,M$_\odot$. Our observations of HIJASS J1021+6842 which were taken during the same observing runs were published separately \cite[(Walter \etal\ 2005)]{walter05}. So far, about a dozen clouds have been detected within a region to the South--East of the triplet (in the vicinity of the dSphs KDG\,64 and BK5N); there were no detections  toward regions around other dSph galaxies in the same group and  in five more ``blind" pointings. 

From a technical perspective, our observations go much deeper than any previous blind surveys done with single dish telescopes in the M\,81 group (\cite[Lo \& Sargent 1979]{lo79}; \cite[Boyce \etal\ 2001]{boyce01}). The barely resolved clouds detected with the $35^{\prime\prime}$ VLA beam fall below the detection limit of single dish telescopes as a result of beam dilution. Our VLA data are also more sensitive by an order of magnitude than similar such surveys done in the Centaurus and Sculptor groups with the ATCA (\cite[de Blok \etal\ 2002]{blok02}\cite[; see also the discussion of the null result with Arecibo by Zwaan \& Briggs 2000)]{zwaan00}.

The HI clouds detected here don't seem to have any optical counterparts. Circumstantial evidence argues in favour of these clouds being debris from the tidal interaction of the galaxies making up the M\,81 triplet, notably the passage of NGC\,3077 sweeping in a prograde fashion around the South of M\,81. In order to better understand these enigmatic clouds further observations (extending the area around BK5N and KDG\,64 covered thus far) are being analysed.

\begin{acknowledgments}
We thank Min Yun for providing us with an electronic version of his 1994 data which were used in Fig.\,\ref{m81}.
\end{acknowledgments}


\begin{thebibliography}{}

\bibitem[Appleton \etal\ (1981)]{appleton81}
     {Appleton, P.N., Davies, R.D. \& Stephenson, R.J.} 1981,
     \textit{MNRAS} 195, 327

\bibitem[Boyce \etal\ (2001)]{boyce01}
     {Boyce, P.J., Minchin, R.F., Kilborn, V.A., Disney, M.J., Lang, R.H., Jordan, C.A., Grossi, M., Lyne, A.G., Cohen, R.J., Morison, I.M. \& Phillipps, S.} 2001,
     \textit{ApJ} (Letters) 560, L127

\bibitem[de Blok \etal\ (2002)]{blok02}
     {de Blok, W.J.G., Zwaan, M.A., Dijkstra, M., Briggs, F.H. \& Freeman, K.C.} 2002,
     \textit{A\&A} 382, 43

\bibitem[Freedman \etal\ (2001)]{freedman01} 
	{Freedman, W.L., \etal} 2001,
	\textit{ApJ} 553, 47 
	
\bibitem[Froebrich \& Meusinger (2000)]{froebrich00}
     {Froebrich, D. \& Meusinger, H.} 2000,
     \textit{A\&AS}145, 229

\bibitem[Huchtmeier \& Skillman (1998)]{huchtmeier98}
     {Huchtmeier, W.K. \& Skillman, E.D.} 1998, 
     \textit{A\&AS} 127, 269
     
\bibitem[Huchtmeier \etal\ (2000)]{huchtmeier00}
     {Huchtmeier, W.K., Karachentsev, I.D., Karachentseva, V.E. \& Ehle, M.} 2000, 
     \textit{A\&AS} 141, 469
          
\bibitem[Karachentsev  \& Kaisin (2007)]{kara07}
     {Karachentsev, I.D. \& Kaisin, S.S.} 2007,
     \textit{AJ}133, 1883

\bibitem[Lo \& Sargent (1979)]{lo79}
     {Lo, K.Y. \& Sargent, W.L.W.} 1979,
     \textit{ApJ} 227, 756

\bibitem[Mateo (1998)]{mateo98}
     {Mateo, M.L. } 1998,
     \textit{ARAA} 36, 435

\bibitem[Pisano \etal\ (2004)]{pisano04} 
	{Pisano, D.J., Barnes, D.G., Gibson, B.K., Staveley--Smith, L., Freeman, K.C. \& 	Kilborn, V.A.} 2004
	\textit{ApJ} (Letter) 610, L17

\bibitem[Pisano \etal\ (2007)]{pisano07} 
	{Pisano, D.J., Barnes, D.G., Gibson, B.K., Staveley--Smith, L., Freeman, K.C. \& 	Kilborn, V.A.} 2007
	\textit{ApJ} 662, 959

\bibitem[Simien \& Prugniel (2002)]{simien02}
     {Simien, F. \& Prugniel, Ph.} 2002,
     \textit{A\&A} 384, 371

\bibitem[Skillman \etal\ (2003)]{skillman03}
     {Skillman, E.D., C\^ot\'e, S. \& Miller, B.W.} 2003,
     \textit{AJ} 125, 593
     
\bibitem[Yun \etal\ (1994)]{yun94}
     {Yun, M.S., Ho, P.T.P. \& Lo, K.Y.} 1994,
     \textit{Nature} 372, 530
     
\bibitem[Yun (1997)]{yun97}
     {Yun, M.S.} 1997, in: J. E. Barnes, and D. B. Sanders (eds.),
     Galaxy Interactions at Low and High Redshift
     \textit{Proceedings of IAU Symposium 186} p.\ 81
     
\bibitem[Yun \etal\ (2000)]{yun00}
     {Yun, M.S., Ho, P.T.P. \& Lo, K.Y.} 2000,
	in: J.G. Mangum \& S.J.E. Radford (eds.),
	Imaging at Radio through Submillimeter Wavelengths, 
	\textit{ASP Conf.\ Proc.} Vol.\ 217, p.\ 374

\bibitem[Walter \etal\ (2005)]{walter05}
     {Walter, F., Skillman, E.D. \&  Brinks, E.} 2005,
     \textit{ApJ} (Letters) 627, L105
     
\bibitem[Zwaan \& Briggs (2000)]{zwaan00}
     {Zwaan, M.A. \& Briggs, F.H.} 2000,
     \textit{ApJ} (Letters) 530, L61

\end{thebibliography}
\end{document}